# SCIENTIFIC REPORTS

**OPEN**

# Current distribution across type II superconducting films: a new vortex-free critical state

E. F. Talantsev[1], A. E. Pantoja[1], W. P. Crump[1] & J. L. Tallon[1,2]

The current distribution across the thickness of a current-carrying rectangular film in the Meissner state was established long ago by the London brothers. The distribution across the width is more complicated but was later shown to be highly non-uniform, diverging at the edges. Accordingly, the standard view for type II superconductors is that vortices enter at the edges and, with increasing current, are driven inwards until they self-annihilate at the centre, causing dissipation. This condition is presumed to define the critical current. However we have shown that, under self-field (no external field), the transport critical current is a London surface current where the surface current density equals the critical field divided by $\lambda$, *across the entire width*. The critical current distribution must therefore be uniform. Here we report studies of the current and field distribution across commercial $YBa_2Cu_3O_7$ conductors and confirm the accepted non-uniform distribution at low current but demonstrate a radical crossover to a uniform distribution at critical current. This crossover ends discontinuously at a singularity and calculations quantitatively confirm these results in detail. The onset of self-field dissipation is, unexpectedly, thermodynamic in character and the implied vortex-free critical state seems to require new physics.

The distribution of a DC transport current across a superconductor is fundamentally different from that across a normal-state conductor because of the inherent issues of Meissner shielding, flux pinning and flux trapping[1,2]. The London brothers[3] showed that, in the Meissner state, currents are confined to the surface within a few $\lambda$, where $\lambda$ is the London penetration depth (see Fig. 1). This describes the current distribution into the superconductor (or for a film, *across the thickness*). Much later Rhoderick and coworkers[4,5] considered thin-film superconductors in the Meissner state and established that the current is distributed non-uniformly *across the width*. To be specific, as shown in Fig. 1, we consider an infinitely long rectangular cross-section superconductor transporting a current, $I$, in the z-direction, with thickness $2b$ extending in the y-direction and width $2a$ extending in the x-direction. Generally, though not necessarily, $a \gg b$. Rhoderick and coworkers showed that the local x-dependent current density, $j(x)$, integrated over the film thickness, is given by:

$$j(x) = \frac{I}{\pi\sqrt{a^2 - x^2}}, \tag{1}$$

and thus diverges at the edges. This scenario has been reproduced by many authors subsequently[6,7]. As a consequence of this divergence it is generally accepted that, at suitably high current, vortices will enter at the edges and migrate inwards under the Lorentz force. If volume pinning is present, the flux front will progress to only a certain depth which increases with further increase in current. Eventually at high enough current the flux front extends to the centre of the conductor and dissipation sets in due to the ongoing ingress and self-annihilation of opposite-sense vortices at the centre. This defines the critical current density $J_c$. Brandt and Indenbom[6] showed that the current distribution across the width of the conductor is progressively modified by this flux penetration and becomes essentially uniform at $J_c$ where, in an ideal conductor, the local current density is $J_c$ everywhere. Even in the absence of pinning, the Rhoderick non-uniform distribution leads to a geometric vortex barrier[8] which prevents vortex entry up to a limiting *local* current density at the edges given by $B_c/(\mu_0\lambda)$. These ideas are well accepted in the field.

[1]Robinson Research Institute, Victoria University of Wellington, P.O. Box 33436, Lower Hutt, 5046, New Zealand. [2]MacDiarmid Institute for Advanced Materials and Nanotechnology, P.O. Box 33436, Lower Hutt, 5046, New Zealand. Correspondence and requests for materials should be addressed to E.F.T. (email: Evgeny.Talantsev@vuw.ac.nz) or J.L.T. (email: Jeff.Tallon@vuw.ac.nz)

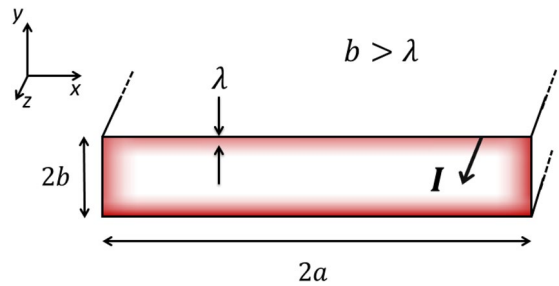

**Figure 1.** The geometry of the rectangular conductors discussed in the text. Current (arrowed) flows in the $z$-direction and the conductor cross-section lies in the $x-y$ plane. It is of width $2a$ in the $x$-direction, extending from $x=-a$ to $x=+a$; and of thickness $2b$ in the $y$-direction extending from $y=-b$ to $y=+b$. In the Meissner state the transport current is just the London surface current confined to within a few $\lambda$ of the surface as illustrated by the shaded regions, and arrows, for the case $b>\lambda$.

Contradicting this, we here focus solely on London transport currents in the Meissner state i.e. in type I superconductors below the critical field, $B_c$, or type II superconductors below the lower critical field, $B_{c1}$. Alternatively, and as it happens equivalently, we focus on self-field transport currents up to the magnitude of the self-field critical current. The link between these two scenarios is not obvious but was established by us in a body of work[9–12] that showed, for both type I and II superconductors, that (i) the self-field transport current is the London surface current in the Meissner state, all the way up to $J_c(\text{sf})$, and (ii) the self-field critical current coincides with the London surface current density reaching the value of $B_c/(\mu_0\lambda)$ for type I or $B_{c1}/(\mu_0\lambda)$ for type II superconductors, *everywhere*. These conclusions were drawn from more than 100 data sets including metals, alloys, cuprates, pnictides, oxides, nitrides, heavy Fermions and borocarbides. They also include type I and II superconductors, of both cylindrical and strip geometry, with aspect ratios ranging from 1 to $3\times 10^6$ and dimensions ranging from just one atomic layer thick through to macroscopic samples of millimetre thickness. Crucially, this second conclusion immediately contradicts the non-uniform current distribution established by Rhoderick. There is no difficulty in the case of round cross-section wires because, by symmetry, the current distribution around the cylindrical surface must necessarily be uniform. But for superconductors deployed with rectangular cross-section we are faced with a clear conflict between our experimental deduction of uniform current distribution and the traditional non-uniform current distribution reported more than 50 years ago for the London-Meissner state.

Here we resolve this conflict. Using a Hall sensor array we measure the local field distribution over the surface of $RBa_2Cu_3O_7$ (RBCO) tapes. Here R is, in general, a mixture of rare-earth elements and/or Y. We show that, at low current, the current distribution is indeed non-uniform and the associated field distribution is precisely as described by Rhoderick in both shape and absolute magnitude. But, as the current increases, the current distribution progressively alters and, at critical current, it has become uniform across the entire width. This is just as we predicted from a scaling analysis of $J_c$ over some eight orders of magnitude in sample dimension[12]. Some of the present data was reported in that paper in confirmation of this prediction[12]. The model presented was *qualitatively* along the following lines. At the edges the local surface current density, $J_s(x)$, first saturates at $B_{c1}/(\mu_0\lambda)$, then with increasing current this domain of saturation progressively moves inwards until, at critical current, $J_s(x)=B_{c1}/(\mu_0\lambda)$ across the entire conductor width. Here, we present a *quantitative* critical-state model that, with no adjustable parameters, accurately captures the totality of the evolution of this current and field distribution. The quantitative agreement is exceptional and our analysis reveals the *true* critical current at which the critical state abruptly extends across the entire conductor width. This *true* critical current is of thermodynamic origins and is smaller than that given by the conventional, but essentially arbitrary, electric field criterion of $1\,\mu V/cm$. For the time being we just discuss the case of type II superconductors but the same ideas apply for type I superconductors with $B_{c1}$ replaced by $B_c$.

We measure the transport critical current and, using a Hall sensor array, the field distribution across the width of RBCO coated-conductor tapes immersed in liquid nitrogen i.e. at 77 K. For each, the measurements are in zero external field and thus confined to self-field transport conditions. Voltage taps for measuring critical current were positioned on the tapes approximately 12 cm apart, with the Hall sensor placed half way between the two voltage taps. Each length of superconducting sample tape used was 15 to 20 cm long between the current supply leads. Voltage measurements were carried out using an Agilent type HP34420A Nano-Voltmeter using a 100 $\mu V$ range, and the transport current was supplied with an Agilent HP-6680A constant-current power supply. Critical current was determined according to the usual $1\,\mu V/cm$ criterion.

The Hall sensor, a seven-element linear array, is a custom type THV-MOD (*Arepoc* s.r.o, Slovakia) operating on an excitation current of 4 mA. The manufacturer's field and temperature sensitivity calibrations were used for each individual sensor. A seven-channel preamplifier with a DC gain of 1300 × was used to amplify the magnetic-field Hall-voltage signals, and data was captured with a National Instruments c-DAQ acquisition system with an integration time of 0.5 s, run under a LabView platform. The absolute accuracy of each sensor was verified using an N38 Nd-Fe-B magnet and a calibrated Hall probe (*Group 3*, type LPT-141), and found to be accurate to <2%. The relative sensitivity of the system is ≤0.02 mT, and each sensor has a specified linearity better than 0.2% up to 1 T. The Hall sensors each have an active area of 0.05 mm², are linearly positioned 1.5 mm apart and lie approximately 0.5 mm above the YBCO film. The sensor array may be centred on the tape or placed off-centre and, further, centred and off-centred measurements overlay quite precisely so as to allow local field measurements across the middle, as well as beyond the edge, of the tape.

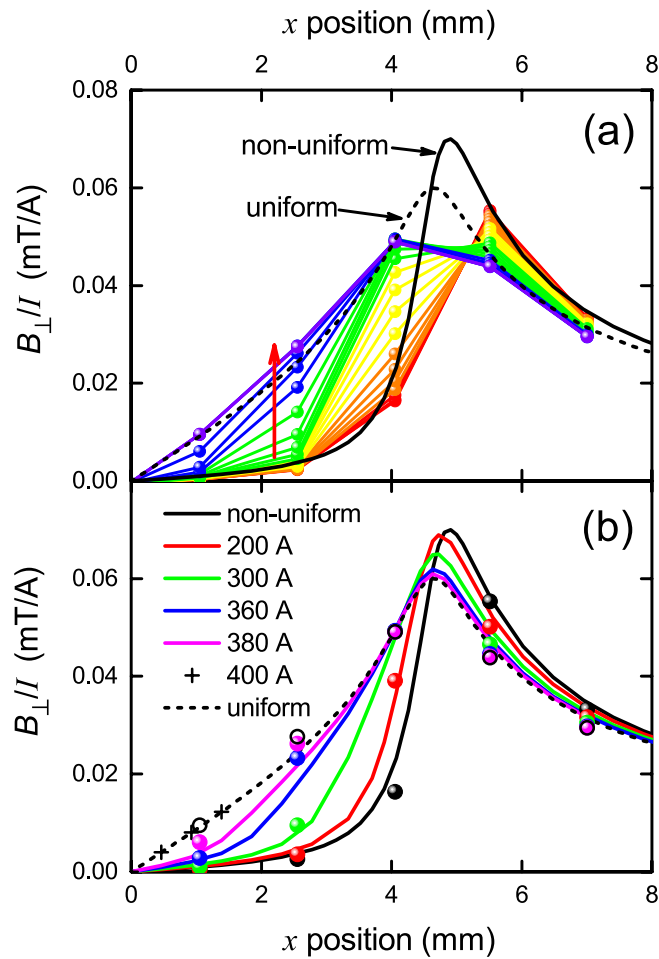

**Figure 2.** (**a**) The *normalised* perpendicular-field distribution, $B_\perp(x, I)$, across a commercial Fujikura RBCO tape measured using a 7 Hall-sensor array for a range of conductor currents from 20 A to 460 A in steps of 20 A (red arrow shows increasing current). The data is plotted as the normalised ratio $B_\perp/I$ so as to show the detailed evolution at lower currents. The sensors are arranged in a line each 1.5 mm apart. The solid black curve is the calculated field distribution (0.5 mm above the tape surface at the Hall sensors) for the Rhoderick and Wilson non-uniform current distribution[4,5] while the dashed black curve is the calculated field distribution for a uniform current distribution. As current increases there is a crossover from non-uniform to uniform current distribution when critical current is attained. (**b**) shows $B_\perp(x, I)$, calculated using the critical state model given by Eq. 4 for $I = 200, 300, 360, 380$ and 400 A. The data points are the measured values for the same currents and are colour coded accordingly. The black data points are for the lowest current, 20 A, which effectively correspond to the Rhoderick and Wilson non-uniform current distribution. There is an excellent match in both shape and absolute magnitude between the theory and the experimental data.

Figure 2(a) shows our measurements (coloured data points) of the local perpendicular magnetic field, $B_\perp(x)$, over the surface of a 10 mm wide commercial Fujikura 2 G RBCO tape. $B_\perp(x)$ is the *y* component of the near-surface field. We show the data just for the positive quadrant of the conductor so that the detailed evolution can be seen better. The conductor in this quadrant extends from $x = 0$ to $x = 5$ mm. Of course by symmetry the field is negative in the left quadrant, $-5 \leq x \leq 0$. We plot the data as $B_\perp/I$ where $I$ is the total transport current which is increased in steps of 20 A from 20 to 460 A (the red arrow indicates increasing current). The measurement at highest current, 460 A, coincides with the self-field transport critical current according to an electric field criterion of $1~\mu V/cm$. Plotted also is the field distribution for a uniform current distribution (dashed curve) and for the Rhoderick and Wilson non-uniform current distribution given by Eq. 1 (solid curve)[4] - more details are provided below. It is evident from the data that, at low current (20–60 A), the field distribution is indeed consistent with the Rhoderick and Wilson non-uniform current distribution. But, crucially, at higher currents the field distribution deviates from this increasingly and has crossed over, at critical current, to a field distribution which is fully consistent with the uniform current distribution which we deduced from our scaling analysis. This correspondence is supported both in the shape and in the absolute magnitudes of the local field, with no fitted parameters. It is important to note that, in the original experimental study used by Rhoderick and Wilson[4,5] on a 20 nm thick Pb strip, the current used was about 1/500$^{th}$ of the critical current. Consistent with our data shown in Fig. 2(a), it is not surprising that at such a low current the non-uniform current distribution was observed. It is our clear prediction that, at 500 times this current density, Pb films or strips will likewise exhibit a uniform current distribution across the conductor width.

We note that the detailed correspondence between calculated and observed fields depends on the accuracy of the positioning of the Hall sensors above the YBCO layer (as noted, approx. 0.5 mm) and on the uniformity of performance of the conductor over its width. However the agreement with the calculated field distributions is impressively good. Many subsequent measurements on tapes from several different manufacturers reinforce in great detail the picture presented in Fig. 1, and for completeness these will be discussed later.

We now describe the calculation of the respective field distributions as well as calculating the detailed crossover between the two extremes, non-uniform and uniform. Let $j(x, I)$ be the local $x$-dependent current density integrated over the film thickness. For example, in the case of the Rhoderick current distribution, $j(x, I)$ is given by Eq. 1 and its shape is independent of the magnitude of $I$. But in general $j(x, I)$ has to be non-linearly $I$-dependent as there exists a current-induced crossover in distribution. Using Ampere's law we can calculate the perpendicular field component at any position $(x_1, y_1)$ arising from a current element, $j(x, I)dx$, at position $x$ on the conductor and integrate over the conductor width to obtain:

$$B_\perp/I = (\mu_0/2\pi) \int_{-a}^{+a} j(x, I) \frac{(x - x_1)}{(x - x_1)^2 + y_1^2} dx. \quad (2)$$

With our geometry $y_1 = 0.5$ mm and we scan $x_1$ from $-10$ mm to $+10$ mm. For the non-uniform Rhoderick charge distribution we use $j(x, I) = j(x)$ given by Eq. 1, while for the uniform distribution we use $j(x) = I/(2a)$, where $I$ is the total current. The result of the numerical integrations for the two limiting cases is shown by the solid and dashed curves in Fig. 2(a). We note that the calculated field distributions differ slightly from those presented by Rhoderick due to the actual offset, $y_1 = 0.5$ mm. For the uniform case Eq. 2 is analytically integrable and we obtain

$$B_\perp/I = (\mu_0/8\pi a) \ln\left[\frac{(a - x_1)^2 + y_1^2}{(a + x_1)^2 + y_1^2}\right]. \quad (3)$$

This is the dashed curve in Fig. 2(b) which, by way of check, precisely matches the numerical calculation, represented in Fig. 2(b) by the three crosses.

Turning now to the $I$-dependent crossover between the two limiting cases, our picture (derived from more than 100 data sets encompassing an exhaustive range of superconductors[10,12]) is that the local critical current is reached when the local surface current density, $J_s(x)$, reaches $B_{c1}/(\mu_0\lambda)$. This commences of course at the edges where at low current the current density is greatest, indeed it is nominally divergent. With increasing current the saturated domain extends deeper to a position $x = \pm d(I)$ where $d < a$. Thus, as current increases, $d$ decreases from the magnitude of $a$ reaching zero at critical current. The problem of current saturation in a thin strip was discussed by Brandt and Indenbom[6] drawing on the earlier work by Norris[13] who developed his equations using conformal mapping from cylindrical to rectangular geometry. The context for Brandt and Indenbom was flux penetration by the ingress and pinning of vortices, very different from the present picture of London-Meissner transport currents confined to within a few $\lambda$ of the surface. We will distinguish between these two models in more detail later.

From Brandt and Indenbom, but confining ourselves to self-field transport currents, the local $x$-dependent current density integrated over the film thickness, $j(x, I)$, is:

$$\begin{aligned} j(x, I) &= \frac{2I_c}{\pi a} \arctan\left(\sqrt{\frac{a^2 - d^2}{d^2 - x^2}}\right), & |x| < d \\ &= I_c/a & d < |x| < a \end{aligned} \quad (4)$$

where $d$ is given by $d(I) = a \times \sqrt{1 - (I/I_c)^2}$ and $I_c$ is the self-field critical current. Thus Eq. 4 may be inserted in Eq. 2 and integrated numerically to obtain the predicted field at the point $(x_1, y_1 = 0.5)$. Figure 2(b) shows the result of this calculation for $I = 20, 200, 300, 360, 380$ and $400$ A (solid curves) with $I_c = 400$ A. The predicted evolution of the field distribution is precisely as observed in experiment. Indeed, we plot in panel (b) the measured data points for these currents using the same colours for identification. We have had to expand the width scale by 8% to allow for the actual YBCO film width as opposed to the physical conductor width, otherwise the field magnitudes are precisely as calculated. The evolution of $B_\perp$ with current, including the inversion that occurs for $|x_1| a$, reproduces the observed behaviour in remarkable detail. Any small differences are entirely attributable to local variations in performance, likely within our model to be due to local variations in London penetration depth, as is observed[14].

It is particularly instructive to consider the current dependence of $B_\perp(x)$. As already pointed out[15] this displays a singular transition significantly below the conventional $I_c$ as defined by the electric field criterion (1 $\mu$V/cm). The data for $B_\perp(x, I)$ is plotted in Fig. 3(a) in steps of 25 amps at each of the 9 locations $x = -5.0, -3.5, -2.0, -0.5, 1.0, 2.5, 4.0, 5.5$ and $7.0$ mm. The values are normalized by dividing by the local $B_\perp$ value at 460 amps, which is the critical current at the 1 $\mu$V/cm threshold. Further, the negative field values for the left-hand quadrant are plotted as positive to better display the detailed evolution. As discussed elsewhere[15], the local perpendicular field is non-linear in $I$ up to 400 amps and then abruptly becomes linear (black dashed line). This location (at 400 A) corresponds exactly to the first observable onset of dissipation, and it occurs at an electric field which is two orders of magnitude below the (arbitrary) 1 $\mu$V/cm criterion. The non-linear behaviour is due to the current-dependent redistribution of current density across the width of the conductor. At 400 amps this redistribution has saturated and there is a singular crossover to linear behaviour at that point. This underscores the fact that, in this self-field regime, the critical current is not merely an arbitrary engineering parameter but, rather, a fundamental thermodynamic parameter defined by a truly singular condition, namely when the critical state reaches the centre of the conductor. Underlying this singular redistribution is the fact that the critical state is achieved when the surface current density reaches $B_{c1}/(\mu_0\lambda)$, everywhere. However, we stress that the abrupt crossover from non-linear to linear behaviour at the *true* critical current is a convenient result of the geometry. In the case of a round cylindrical conductor there will be no such crossover because the current distribution around the circumference is of course by symmetry uniform at all times (in the absence of an external magnetic field). The common defining feature in either case is the onset of the surface critical state at $J_s = B_{c1}/(\mu_0\lambda)$. For our rectangular conductors this results in an abrupt crossover from non-linear to linear behaviour whereas for cylindrical conductors this is necessarily absent. Indeed, by Ampere's law the azimuthal field for a cylindrical conductor is always linear in current. What our rectangular film data most crucially reveals is this *singular* nature of the surface critical state which we have noted, therefore underscoring its fundamentally thermodynamic character. For a cylindrical conductor the onset of the critical state will also be singular but the geometry precludes observing this as a change from non-linear to linear behaviour.

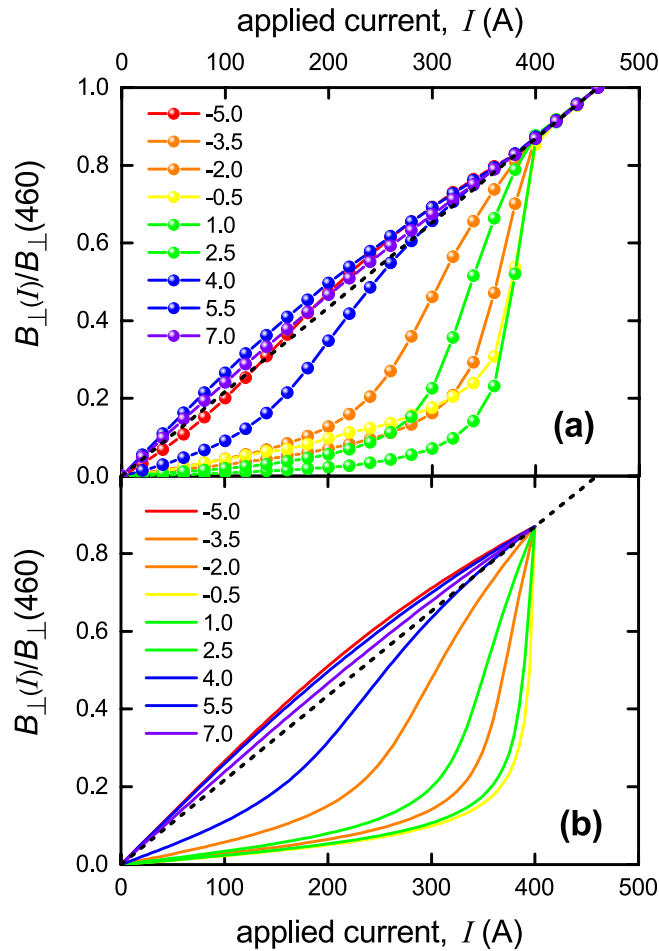

**Figure 3.** (**a**) The current dependence of the *normalized* local perpendicular field, $B_\perp(x)$, at nine locations across a Fujikura commercial RBCO tape, $x = -5.0, -3.5, -2.0, -0.5, +1.0, +2.5, +4.0, +5.5$ and $+7.0$ mm (solid coloured curves and data points). The data is normalized by the respective values of $B_\perp(x)$ at an applied current of 460 A which is the critical current according to a 1 $\mu$V/cm criterion. The non-linear behaviour reveals a singular crossover to linear behaviour (dashed line) at the *true* critical current of 400 A where the entire conductor width is then in the critical state. The locations near the centre, $-2.5 \leq x \leq +2.5$, most clearly show the singular discontinuity. (**b**) Shows the values of $B_\perp(x, I)/B_\perp(x, I = 460\,\text{A})$ calculated using Eqs 2 and 4 with no adjustable parameters. The correspondence with the experimental data is excellent with only small deviations near the centre of the conductor where the absolute values of $B_\perp$ are very small and any deviation is amplified by the large normalization factor there.

This is further illustrated in Fig. 3(b) where we use the above eqs (2 and 4) to calculate the current dependence of $B_\perp(x)$. These calculations almost perfectly reproduce the experimental data and we recall that there are no adjustable parameters here. The only small differences occur near the centre of the tape where the absolute field is very small and any deviations are greatly amplified by rescaling by $B_\perp(x,460\,A)$ when $x$ is small. Again the singular character of the crossover to linear behaviour (black dashed line) is very evident for the locations closer to the tape centre. The remaining puzzle that this uncovers so lucidly is just why the local critical state should be triggered precisely when the local surface current density reaches $J_s(x,I) = B_{c1}/(\mu_0\lambda)$. The physics necessarily does not involve vortices and, in our view, is likely to be new. The fact that identical behaviour is found for type I superconductors (but with $B_{c1}$ replaced by $B_c$) only serves to underscore this point. For type I superconductors the self-field critical state is associated with the onset of depairing and it is quite likely to be the same for type II superconductors.

| Manufacturer | Fujikura | Superpower | Theva |
|---|---|---|---|
| Manufacturer's wire ID | FYSC-S10 10-0025-01 | SCS12050-AP M4-382-5 | TPL1100, ID 170468 |
| Nominal width (mm) | 10 | 12 | 12 |
| SC film thickness ($\mu$m) | 2.3 | 1.5 | 2.6 |
| Composition (RE) | Gd | Y,Gd | Gd |
| $J_c$(77.4 K) (MA/cm$^2$) | 2.0 | 2.9 | 1.2 |
| $n$-value | 35 | 31 | 40 |
| Technology | IBAD/PLD | IBAD/MOCVD | ISD/E-beam co-evaporation |
| Substrate | Hastelloy | Hastelloy C276 | Hastelloy C276 |
| Architecture | Al$_2$O$_3$/Y$_2$O$_3$/MgO/CeO$_2$/GdBCO/Ag | MgO/MgO/MgO/LnMnO$_3$/(Y,Gd)BCO+BaZrO$_3$/Ag/Cu | MgO/GdBCO/Ag |

**Table 1.** Specifications for the RBCO 2 G tapes investigated in this work.

We repeated these measurements with other commercial RBCO coated-conductor tapes and found essentially identical behaviour. Specifications for these tapes are given in Table 1. Figure 4 shows a comparison of measurements between (a) the above-described 10 mm wide Fujikura tape, (b) a 12 mm wide Superpower tape and (c) a non-certified developmental Theva tape. In the first two cases ((a) and (b)) the seven Hall sensor array is approximately centred on the middle of the tapes, while the third is off-centred. In panel (a) the disparity at the extreme left between data and models reflects a small asymmetry in alignment or in conductor performance. In panel (b) all measured fields are slightly higher than the calculated curves but an almost perfect match is achieved if the effective RBCO film width within the tape is 11.7 mm rather than the nominal 12 mm. In panel (c) the field at the edges is consistent with an effective 9 mm wide conductor despite the nominal 12 mm width. This suggests some damage at the edges of the film and we stress that the tape was not certified and our results should not reflect on certified product. We note that our measurements are an exacting indicator of the uniformity of quality over the conductor width. Leaving aside these minor variations it is evident that all these tapes exhibit essentially identical behaviour in crossing over from a non-uniform Rhoderick current distribution at low current to a uniform distribution at high current. Combined with our previous conclusions derived from our scaling analysis we conclude that this crossover is a universal property of self-field currents in superconductors.

An obvious superficial inference might be that we are simply observing the self-field evolution of current distribution associated with vortex entry at the film edges as described by Brandt and Indenbom, and we address this now. The idea in this picture is that, when flux has partly penetrated, the flux lines rearrange such that $j(x) \leq j_c$ at all points in the sample. A critical state with maximal $j(x) = j_c$ is established near the edges of the film and a smaller $j(x)$ flows over the entire remaining width of the strip so as to shield the central flux-free region. With increasing current the flux-free region narrows and the critical state moves in towards the centre. For convenience the local $j_c$ was assumed to be uniform and its magnitude is set by the pinning parameters relevant to the microstructure of that particular sample. This model is reproduced quantitatively in our experiments so why should it not physically describe what is going on? There are three key reasons as to why not:

(i). the self-field $J_c$ which we have reported (now from more than 100 data sets) is found to be independent of pinning[9,12]. Pinning strategies that substantially raise the *in-field* $J_c$ do not appear to alter the *self-field* $J_c$. It is difficult to see how this could arise under the Brandt and Indenbom model.

(ii). by studying these >100 data sets (embracing sample dimensions ranging from single-atomic layer thickness to mm-size) we found $J_c$ always satisfies the scaling relation $J_c = B_{c1}/(\mu_0\lambda) \times (\lambda/b)\tanh(b/\lambda)$. (For type I replace $B_{c1}$ by $B_c$). This means that for $b \gg \lambda$ the critical current density falls as $(1/b)$ and this in turn arises because, at self-field $J_c$, all these samples are in the London-Meissner state. The $(1/b)$ fall-off occurs because the London transport current is confined to the surfaces with "dead-space" between. This contrasts the Brandt and Indenbom picture where the transport current is distributed across the entire sample thickness. Only when $b \gg \lambda$ is the London current distributed over the full sample thickness and this is the reason why the scaling behaviour then crosses over to the $b$-independent behaviour, $J_c = B_{c1}/(\mu_0\lambda)$, for all samples. (To check this, note that $(\lambda/b)\tanh(b/\lambda) \to 1$ when $b \gg \lambda$, and $(\lambda/b)\tanh(b/\lambda) \to \lambda/b$ when $b \gg \lambda$).

(iii). the fact that the self-field $J_c$ scaling is the same for both type I and type II superconductors would suggest that a similar mechanism is operative in both cases. The Brandt and Indenbom vortex mechanism cannot be applicable to type I superconductors which remain in the London-Meissner state right up to self-field $J_c$.

Figure 5 summarises the $I$-dependent evolution of current and field distribution as calculated using Eq. 4. There exists a self-field critical state for $J \leq J_c$ where the local surface current density, $J_s$, saturates at a value given by the critical field divided by $\lambda$ arising from a mechanism(s) yet to be determined. As shown in Fig. 5(b) this critical state begins at the edges and, with increasing current, moves inwards until the entire film is in the critical state with a uniform current distribution across its width. The completion of this uniform critical state occurs in singular fashion with $(\partial B_\perp/\partial I)|_{|x|\ll a}$ diverging at the *true* critical current - which, in all cases, is less than the critical current given by the essentially arbitrary 1 $\mu$V/cm criterion. As shown in Fig. 5(a) the associated field profile evolves from nearly flat across the centre at low current to nearly linear at critical current. The calculation of these profiles is fully corroborated by the present measurements.

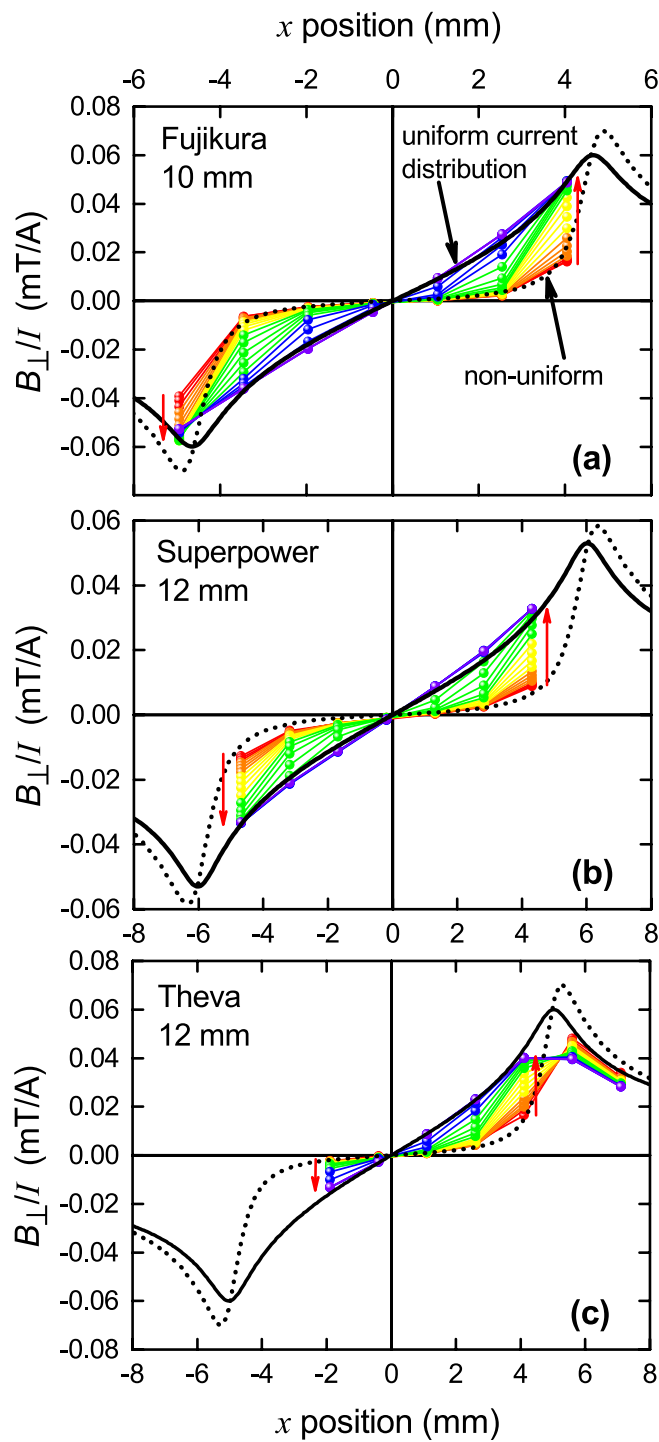

**Figure 4.** The current-normalised perpendicular-field distribution, $B_\perp(x)/I$, across commercial RBCO tapes measured using a 7 Hall-sensor array for (**a**) a 10 mm wide Fujikura conductor with current, $I$, increasing from 20 A to 460 A in steps of 20 A; (**b**) a 12 mm wide Superpower conductor with current increasing from 25 A to 425 A in steps of 25 A; and (**c**) a non-certified 12 mm wide Theva tape with current increasing in steps of 20 A to 380 A. Increasing current is indicated by the red arrows. The sensors are arranged in a line 1.5 mm apart. The black dotted curve is the calculated field distribution (0.5 mm above the tape surface at the Hall sensors) for the Rhoderick and Wilson non-uniform current distribution[4] while the solid black curve is the calculated field distribution for a uniform current distribution.

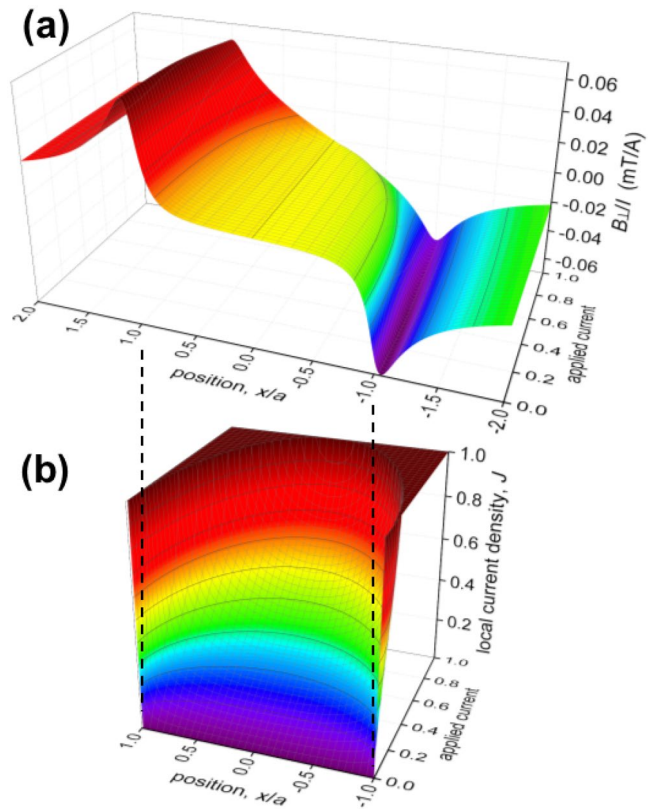

**Figure 5.** (**a**) The distribution of the current-normalised perpendicular field, $B_\perp(x, I)/I$, and (**b**) the distribution of normalized current density, $J(x, I)$, across the width of a rectangular film plotted versus the normalized applied current as calculated using Eq. 4.

In summary we have carried out measurements of the field distribution across the width of several commercial coated-conductor RBCO tapes under self-field transport conditions. We find that, at low current, the field distribution is consistent with the expected non-uniform current distribution advanced by Rhoderick and co-workers (see Eq. 1). But, with increasing current, the profile evolves into a uniform distribution at critical current. This evolution is consistent with the occurrence of a critical state when the surface current density, $J_s$, reaches $B_{c1}/(\mu_0 \lambda)$. By imposing the boundary condition that $J_s$ never exceeds this critical value we successfully account for the detailed quantitative evolution of the current and field profiles with no adjustable parameters. From the London equations this critical state corresponds to a current density averaged over the thickness given by $J = B_{c1}/(\mu_0 b) \times tanh(b/\lambda)$[12]. With increasing applied current the saturated region progressively moves in from the edges and eventually, at self-field critical current, the local current density is $J_c = B_{c1}/(\mu_0 b) \times tanh(b/\lambda)$, uniformly across the entire width. The onset of dissipation at the self-field critical current is *singular* and therefore of thermodynamic origin. It is not subject to an arbitrary electric-field criterion. The associated field distribution evolves from nearly flat across the centre at low current to nearly linear at critical current. This picture differs from previous vortex entry/pinning models in that the self-field transport current is in fact the London current confined to a few $\lambda$ of the surface. A remaining task is to identify the origins of this self-field critical state[12].

EFT and JLT separately thank the Marsden Fund of New Zealand for financial support (EFT: grant number VUW1608, JLT: grant number VUW1322). We also thank Dr Rod Badcock, Robinson Research Institute, Victoria University of Wellington, for making the Hall probe facility available for these experiments.

### Author Contributions

E.F.T. and J.L.T. jointly conceived this study; E.F.T. and A.E.P. set up the experiment; E.F.T. performed the experiment and collected the data; E.F.T., W.P.C. and J.L.T. carried out the modelling and calculations; J.L.T. drafted the manuscript which was revised and edited by all parties.

### Additional Information

**Competing Interests:** The authors declare that they have no competing interests.

**Publisher's note:** Springer Nature remains neutral with regard to jurisdictional claims in published maps and institutional affiliations.